\pdfoutput=1
\documentclass[preprint,12pt]{elsarticle}
\usepackage{graphicx}
\usepackage{amsmath}
\usepackage{hyperref}
\usepackage{booktabs}
\usepackage{multirow}

\usepackage{geometry} 
\journal{Computer Speech \& Language}

\begin{document}

\begin{frontmatter}

\title{Split and Conquer Partial Deepfake Speech}

\author[inst1]{Inbal Rimon\corref{cor1}}
\ead{inbalri@post.bgu.ac.il}
\author[inst2]{Oren Gal}
\author[inst1]{Haim Permuter}

\cortext[cor1]{Corresponding author}

\address[inst1]{Ben Gurion University, Be'er Sheva, Israel}
\address[inst2]{University of Haifa, Haifa, Israel}

\begin{abstract}
Partial deepfake speech detection requires identifying manipulated
regions that may occur within short temporal portions of an
otherwise bona fide utterance, making the task particularly challenging
for conventional utterance-level classifiers. We propose a
split-and-conquer framework that decomposes the problem into two stages: boundary detection and segment-level classification. A dedicated boundary detector first identifies temporal transition points, allowing the audio signal to be divided into segments that are expected to contain acoustically consistent content. Each
resulting segment is then evaluated independently to determine whether it corresponds to bona fide or fake speech.

This formulation simplifies the learning objective by explicitly
separating temporal localization from authenticity assessment, allowing
each component to focus on a well-defined task. To further improve
robustness, we introduce a reflection-based multi-length training
strategy that converts variable-duration segments into several fixed
input lengths, producing diverse feature-space representations. Each stage is trained using multiple configurations with different feature extractors and augmentation strategies, and their complementary predictions are fused to obtain improved final models.

Experiments on the PartialSpoof benchmark demonstrate state-of-the-art
performance across multiple temporal resolutions as well as at the
utterance level, with substantial improvements in the accurate detection
and localization of spoofed regions. In addition, the proposed method achieves state-of-the-art performance on the Half-Truth dataset, further confirming the robustness and generalization capability of the framework.

\end{abstract}


\end{frontmatter}

\section{Introduction}

Recent advances in neural text-to-speech (TTS), voice conversion (VC), and diffusion-based generators have made it straightforward to synthesize highly natural-sounding speech. While these technologies enable a wide range of beneficial applications, they also facilitate audio deepfakes that can undermine trust in spoken communication and threaten security-critical systems such as automatic speaker verification (ASV) and voice-based authentication. In response, a substantial body of work on speech anti-spoofing and audio deepfake detection has emerged, supported and systematized by the ASVspoof challenge series \cite{asvspoof2019,asvspoof2021,asvspoof5} and has been reviewed in depth in recent surveys on audio deepfake detection and anti-spoofing \cite{yi2023_audio_survey,zhang2025_audio_survey}. However, the majority of existing countermeasures assume that each utterance is either fully bona fide or fully spoofed and therefore operate at the utterance level.

A more realistic and increasingly emphasized threat model is partial manipulation, in which only a subset of an utterance is replaced or in-painted with fake speech while the rest remains genuine \cite{zhang2021initial,yi2021_halftruth}. Partial deepfakes are attractive to attackers because they allow key phrases (for example, names, amounts, or authorization statements) to be modified while preserving most of the original speech, thereby improving plausibility. From a detection perspective, they are particularly challenging: boundaries between bona fide and spoofed regions can be ambiguous, local artifacts may be subtle and short-lived, and global utterance-level cues can be diluted by surrounding bona fide speech.

\subsection{Full-utterance deepfake detection}

Work on full-utterance deepfake detection typically casts the problem as binary classification at the utterance level: given an input recording, the system outputs a single bona fide or spoof label. Early countermeasures employed hand-crafted spectral features such as LFCC, MFCC, or CQCC combined with shallow classifiers \cite{todisco2017_cqcc,kinnunen2017_asvspoof_cm}, while later systems moved to deep convolutional or residual networks operating on spectrogram-like representations \cite{lavrentyeva2019_stc,wang2021_asvspoof_cnn}. More recent approaches adopt end-to-end models with learning-based features, including x-vector embeddings and self-supervised speech encoders such as wav2vec 2.0 and HuBERT, often coupled with pooling and attention mechanisms \cite{Li2023VoiceDeepfakeHuBERT,Martindonas2022Wav2Vec2ADD}. 

\subsection{Partially deepfaked audio detection}

Research on partial deepfake detection has mainly followed two directions: frame-based models and utterance-level models with weak localization.

Frame-based approaches typically split the signal into short, fixed-length units and classify each unit independently \cite{zhang2021multi}. Although conceptually simple, this strategy can fragment contextual information and provides limited evidence for each decision, and its effectiveness tends to degrade as the temporal resolution is reduced to shorter frames, where each unit contains only a small portion of the acoustic context.

    A second line of work addresses frames via full-sequence labeling, training models to assign a label to every frame across the utterance \cite{cai2023_waveform_boundary}. Other methods aim to detect both boundaries and authenticity jointly, for example by predicting boundary maps and spoof labels in a single network, or by incorporating boundary-aware attention mechanisms \cite{zhong2024_bam,Cai2024BoundaryDeepfake}. These designs preserve global context, but also introduce practical and statistical challenges: long sequences strain memory and optimization; boundary ambiguity produces label noise near transitions; and a single shared representation is required to support both boundary detection and authenticity classification under a unified objective. This tight coupling can make learning less stable and can limit the effective temporal resolution of each subtask.


\subsection{Benchmarks and datasets}

Progress in partial deepfake detection has been enabled by datasets with explicit temporal annotations. The PartialSpoof (PS) database has become the primary benchmark, providing both utterance-level and frame-level labels at multiple temporal resolutions \cite{zhang2021initial}. The Half-Truth (HAD) dataset emphasizes fine-grained word or phrase manipulations and offers accurate boundaries for temporal localization studies \cite{yi2021_halftruth}. 
These datasets support controlled analysis of localization accuracy, robustness to short edits, and generalization toward more realistic settings.

\subsection{Our contributions}
We introduce a split-and-conquer framework for partial deepfake speech detection that decomposes the task into two dedicated stages: boundary detection and segment-level classification. This design separates temporal localization from authenticity assessment, reducing learning complexity and allowing each component to be trained
under a well-defined objective. The main contributions of this work are as follows.
\begin{itemize}
    \item We design and train a dedicated boundary detector that predicts
    transition points between bona fide and spoofed regions within partially manipulated utterances. Within the proposed pipeline, the boundary detector partitions the audio into candidate spoof-uniform segments, allowing the downstream classifier to operate on temporally consistent units.
    By focusing training solely on transition
    localization, the detector learns robust temporal cues and reduces sensitivity to small timing misalignments that typically affect joint localization–classification approaches.

    \item We propose a boundary-guided segment classification stage in which predicted boundaries define uniform candidate segments that are then processed by a segment-level binary classifier trained on ground-truth–derived segments. This formulation allows us to use a simple classifier on uniform segments and mitigates label noise that occurs when analysis windows span both bona fide and spoofed speech.

    \item We develop a feature-space augmentation strategy based on reflected fixed-length inputs, in which variable-length segments are mapped to multiple target durations without introducing new content. This process yields distinct feature-space representations of the same underlying signal, organized differently in time. When combined through fusion across input lengths and feature extractors, these complementary views lead to improved robustness and overall detection performance.
\end{itemize}

\section{Proposed Method}

\begin{figure}[h!]
\centering
\includegraphics[width=1.0\textwidth]{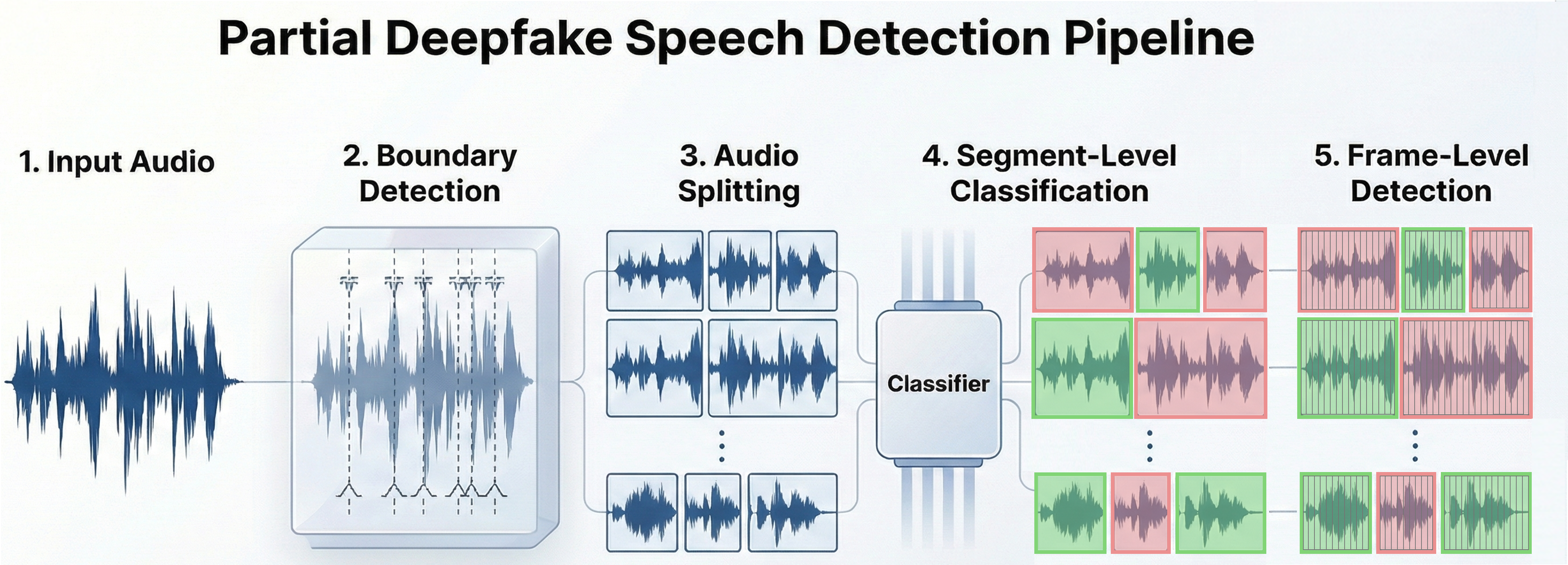}
\caption{Overview of the proposed partial deepfake speech detection pipeline. 
(1) \textbf{Input Audio:} a full utterance containing both bona fide and manipulated regions. 
(2) \textbf{Boundary Detection:} a frame-level model predicts transition points between acoustic regions. 
(3) \textbf{Audio Splitting:} the signal is partitioned into segments according to the detected boundaries, producing candidate spoof-uniform regions. 
(4) \textbf{Segment-Level Classification:} each segment is independently evaluated by a classifier that assigns an authenticity score. 
(5) \textbf{Frame-Level Detection:} segment predictions are projected back onto the temporal axis to obtain fine-grained localization of manipulated speech at frame resolution.}
\label{fig:model}
\end{figure}

We formulate partial deepfake detection as a joint segmentation and classification problem. The goal is to determine the temporal extent of manipulated regions and assign an authenticity label to each region, thereby partitioning an utterance into bona fide and spoof segments. To this end, we employ a two stage pipeline consisting of boundary estimation and segment classification.

In the first stage, a dedicated boundary detector operates on the full utterance and estimates a frame-wise boundary indicator sequence at a 20\,ms temporal resolution. Rather than predicting authenticity labels for individual frames, the model is trained to detect transition points between adjacent temporal regions, where the underlying class changes from bona fide to spoof or vice versa. Each indicator therefore represents the likelihood that a class transition occurs between two consecutive frames. Formulating the task as transition detection instead of frame-level classification directs the model to boundary cues marking manipulated regions while still using global utterance context. This formulation concentrates supervision on a single, well-defined localization objective. It also avoids the ambiguity that arises when analysis windows span both bona fide and spoofed content.

In the second stage, the predicted boundary indicators are converted into candidate spoof-uniform segments. These segments are passed to a segment classifier that outputs an authentication score for each segment. The segment classifier is trained on uniform segments derived from ground truth annotations so that each training example corresponds directly to one decision unit at inference time. Segment level predictions are then mapped back to the frame resolution by assigning each the score of the segment to which it belongs, thus linking local classification decisions to the global segmentation produced by the boundary detector.
The overall processing pipeline is illustrated in Figure~\ref{fig:model}, which summarizes the proposed framework from input audio to frame-level manipulation detection.
Supplementary materials and method highlights are publicly available in the project repository \footnote{\href{https://github.com/InbalRim/SplitAndConquer}{https://github.com/InbalRim/SplitAndConquer}}.

\subsection{Model}

The proposed framework employs two dedicated models for boundary localization and segment-level authenticity classification that address distinct sub-tasks while sharing a common architectural backbone. Decoupling the two tasks simplifies the training process by allowing each model to focus on a single, well-defined objective, which improves optimization stability and interpretability. This design further enables independent evaluation and monitoring of each component, and ensures that performance differences can be attributed to task formulation and supervision.

Both the boundary detector and the segment classifier follow the same high-level structure, consisting of a feature extraction front end followed by a ResNet34~\cite{he2016_resnet} classification head. Using a unified architecture across stages isolates the impact of input structure and training objectives, while maintaining comparable model capacity for both localization and classification.

We consider three feature extractors with complementary representational properties. For contextualized speech representations, we adopt  the wav2vec~2.0 XLSR models pretrained on 53 and 128 languages, denoted as XLSR53 and XLSR128, respectively~\cite{conneau2020_xlsr,babu2022_xlsr}. These self-supervised encoders are trained on large-scale multilingual speech corpora and have been shown to capture robust phonetic and prosodic information across diverse languages and recording conditions. Such properties are particularly advantageous for partial deepfake detection, where manipulated regions may be short, subtle, and subject to substantial cross-corpus variability.

As a non-trainable baseline, we additionally employ a log-magnitude spectrogram front end, following our previous work on full deepfake detection~\cite{rimon2022study}. Although spectrogram-based features lack the inductive biases introduced by self-supervised pretraining, they remain simple, transparent, and widely used in classical anti-spoofing systems. Importantly, their inclusion provides architectural and representational diversity when combined with XLSR-based models in fusion, which we find to be beneficial for boundary localization.

For boundary detection, the model processes the full utterance and outputs a frame-wise boundary probability sequence defined at a 20\,ms temporal resolution. The temporal dimension is preserved throughout the network by appropriately configuring the stride and pooling operations in the convolutional layers, ensuring that each output score remains aligned with its corresponding input frame and enabling precise localization of transitions between bona fide and spoofed regions.

For segment-level classification, the input is expected to be a single spoof-uniform segment extracted from the utterance, and the output is a binary authenticity score for that segment. Architecturally, this classifier resembles a conventional utterance-level deepfake detector; however, it is trained on short, ground-truth-derived segments rather than full utterances. Aligning training samples with inference-time decision units substantially reduces label noise near manipulation boundaries and encourages the learned representation to focus on segment-level cues that are directly relevant to partial deepfake detection.

\subsection{Augmentations}

We employ two data augmentation strategies introduced in our previous work \cite{rimon2025unmasking}. The first augmentation, MaskedSpec (MS), operates on the utterance by applying random frequency-band masking in the time–frequency domain and then reconstructing the waveform. The second, MaskedFeature (MF), operates directly in the latent representation space by applying Gaussian masking to contiguous regions of the feature sequence. Both are applied during training and are intended to improve robustness to spectral distortions and representation-level perturbations, encouraging the models to rely on more global and invariant cues rather than narrow, dataset-specific artifacts.

For the segment classifier, we further introduce a feature-space augmentation motivated by variable segment durations. Because ground-truth segments exhibit a wide range of lengths, each segment is mapped to a fixed input duration by reflection. Concretely, the segment waveform is time-symmetrically reflected and repeated until the target length is reached. This procedure preserves the original content while ensuring that concatenation points remain smooth, thereby avoiding the unnatural discontinuities and spectral artifacts that often arise with zero padding or naive tiling.
Following the recommended input duration in the XLSR literature, we primarily train with a 4 second (4sec) fixed input length for XLSR-based models. In addition, we train variants with 2sec and 1sec fixed lengths to explore the effect of shorter contexts and to reduce computational cost. While the 4sec configuration yields the strongest single-model performance, fusing the three classifiers that differ only in fixed input length consistently improves results. This behaviour suggests that different fixed lengths induce distinct feature-space embeddings of the same underlying segment, exposing the classifier to complementary temporal organizations and receptive-field patterns. We therefore treat variable fixed-length training as an implicit feature-space augmentation and incorporate multi-length fusion, together with fusion across feature extractors and augmentation schemes, in our final system.

\subsection{Boundary Extraction from Prediction Scores}
\label{sec:boundary_thresholding}

The boundary detection model produces a frame-wise sequence of boundary
scores
\begin{equation}
\mathbf{s} = \{s_t\}_{t=1}^{T},
\end{equation}
where $s_t$ denotes the predicted likelihood of a transition between
frames $t$ and $t+1$. Let $N$ denote the total number of frames in the
utterance. Since scores are defined for pairs of consecutive frames, the
sequence length satisfies $T = N - 1$. The objective is to convert this
continuous score sequence into a binary boundary indicator sequence
\begin{equation}
b_t \in \{0,1\},
\end{equation}
indicating whether a boundary occurs at time index $t$.

A direct approach applies a fixed global threshold $\tau_g$,
\begin{equation}
b_t =
\begin{cases}
1, & s_t \ge \tau_g, \\
0, & \text{otherwise},
\end{cases}
\end{equation}
however, variability in score distributions across utterances makes a
single global threshold insufficiently robust. Differences in utterance
duration, speaking style, and acoustic conditions may alter the scale
and dynamic range of the predicted scores, causing a fixed threshold to
either suppress true boundaries or produce excessive false detections.

To address this limitation, we adopt an utterance-dependent
histogram-based thresholding procedure. For each utterance, the boundary
scores are summarized using a histogram with a fixed number of bins,
\begin{equation}
h(i) = \sum_{t=1}^{T} \mathbf{1}\!\left[s_t \in \mathcal{B}_i\right],
\qquad i = 1,\dots,M,
\end{equation}
where $\mathcal{B}_i$ denotes the $i$-th score bin,
$\mathbf{1}[\cdot]$ is the indicator function. Let $\{\beta_i\}_{i=1}^{M+1}$ denote the corresponding bin
edges.

The threshold is then determined from the minimum point of the
histogram, corresponding to the least populated bin,
\begin{equation}
i^{*} = \arg\min_{i} h(i),
\qquad
\tau^{*} = \beta_{i^{*}+1}.
\end{equation}
This bin represents the minimum of the histogram and typically lies
between the dense region of low boundary scores and the sparser region
of higher scores associated with transition events. Selecting the
threshold at the upper edge of this bin provides a conservative
separation between non-boundary and boundary candidates while adapting
to the score distribution of each utterance.

The final binary boundary sequence is obtained by applying the
utterance-specific threshold,
\begin{equation}
b_t =
\begin{cases}
1, & s_t \ge \tau^{*}, \\
0, & \text{otherwise}.
\end{cases}
\end{equation}

This histogram-based procedure converts the continuous boundary scores into discrete transition points while remaining robust to variations in score scale and distribution across utterances.

\section{Results}
In this section, we report results for each stage of the proposed framework and for the full end-to-end pipeline. Unless otherwise stated, all experiments are conducted on the PartialSpoof dataset with metrics computed from a 20\,ms reference segmentation.

\subsection{Experimental Setup}

All models are trained on the training partition of the PartialSpoof dataset, using a batch size of 16, on a single NVIDIA GeForce RTX 3090 GPU for 20 epochs. The development set of PartialSpoof is used for epoch level assessment and model selection. We use the AdamW optimizer with a learning rate and weight decay of \(1\times10^{-6}\). Each input waveform, whether a full utterance or an extracted segment, is pre-emphasis before feature extraction.

Training labels for all tasks are derived from the ground truth segmentation at a 20 ms resolution. For the boundary detection model, we construct a binary label for each pair of consecutive 20 ms frames: the label is set to \(1\) if the two frames belong to different ground truth segments (real versus spoof) and \(0\) otherwise. In other words, a positive label is assigned whenever two consecutive frames are not uniform in terms of spoof versus real content.

For the segment classifier, training segments are defined as maximally contiguous runs of 20 ms frames that share the same ground truth label. Each such segment is treated as a single training example and inherits the common real or spoof label.

For evaluation, we distinguish between the standalone segment classifier and the full pipeline. For the standalone classifier, boundaries are derived directly from the ground truth segmentation, so performance reflects the classifier under ideal boundary information. For the full pipeline, metrics are computed using boundaries predicted by the boundary detector, providing the realistic estimate of end to end system behaviour.

\textit{Frame-level EER (F-EER):}
We represent the reference segmentation at a 20 ms frame resolution and assign a ground truth label to each frame. During evaluation, each frame receives a scalar score from the model, and the EER is computed across all frames.

\textit{Segment-level EER (S-EER):}
Segment labels are derived from the segmentation as temporally contiguous regions with identical labels. Segment scores are obtained by aggregating the model output over the frames that belong to each segment (for example by averaging). Each segment is then treated as an independent example, and the EER is computed across all segments.

\textit{Per-Utterance EER (U-EER):}
To reduce the dominance of very long utterances, we also report utterance averaged metrics. For each utterance, we compute an equal error rate based on either frame level or segment level scores restricted to that utterance, and then average these per utterance EER values over the evaluation set. This yields utterance averaged frame level EER and utterance averaged segment level EER.

\subsection{Step 1: Boundary Detection}
\label{sec:boundary_results}

\begin{table}[h]
\centering

\begin{tabular}{lc|cc|cc}
\toprule
\multirow{2}{*}{Feature extractor} & \multirow{2}{*}{Augmentation} & \multicolumn{2}{c}{PartialSpoof F-EER (\%)} & \multicolumn{2}{|c}{HAD F-EER (\%)} \\
& & DEV & EVAL& DEV & EVAL\\
\midrule
\multirow{3}{*}{XLSR128} 
  & None & 2.65 & 3.85 & 0.12 & 0.15\\
  & MS   & \textbf{2.32} & \textbf{3.47}  & 0.07 & \textbf{0.09} \\
  & MF   & 2.35 & 4.98 & 0.12 & 0.14 \\
\midrule
\multirow{3}{*}{XLSR53} 
  & None & 2.86 & 4.19 &  0.07 & 0.14 \\
  & MS   & 2.51 & 3.71 & \textbf{0.05} & \textbf{0.09} \\
  & MF   & 2.37 & 5.16  & 0.11 & 0.11 \\
\midrule
\multirow{3}{*}{Spectrogram} 
  & None & 3.48 & 5.32 & 0.30 & 0.45 \\
  & MS   & 3.26 & 5.18 & 0.12 & 0.25 \\
  & MF   & 3.50 & 5.24 & 0.17 & 0.27 \\

\bottomrule
\end{tabular}
\caption{Boundary detection performance measured as frame-level EER (\%) on the PartialSpoof and HAD development (DEV) and evaluation (EVAL) sets at 20\,ms temporal resolution for different feature extractors and augmentation schemes.}
\label{tab:boundary_single}
\end{table}

We first analyze the boundary detection independently in order to
assess how reliably temporal transition cues can be identified at a
20\,ms frame resolution. Different feature extractors and augmentation
strategies are compared on both development and evaluation partitions of
PartialSpoof and HAD datasets to examine robustness and generalization behavior. The
results indicate that boundary-related information can be captured
consistently across representations, while frequency-domain masking
provides a reliable improvement in generalization performance.
Table~\ref{tab:boundary_single} summarizes boundary detection performance across different feature extractors and augmentation schemes on both the development and evaluation sets. Across feature extractors, MaskedSpec (MS) consistently improves performance compared to training without augmentation, while MaskedFeature (MF) generally leads to higher evaluation EERs despite competitive development-set results. This discrepancy suggests that MF may encourage representations that fit the training distribution well but generalize less effectively when evaluated under unseen conditions.

A comparison between development and evaluation results further reveals consistent relative ordering among configurations, indicating stable optimization behavior across datasets. At the same time, the larger performance spread observed on the evaluation set suggests that generalization, rather than training convergence, constitutes the primary source of variation between models.

\begin{table}[h]
\centering
\begin{tabular}{lc}
\toprule
Fusion Configuration & F-EER (\%) \\
\midrule
PartialSpoof, EVAL, All & 2.49 \\
HAD, EVAL, All & 0.08\\
\bottomrule
\end{tabular}
\caption{Frame-level EER (F-EER) of the fused boundary detector evaluated on the PartialSpoof and HAD evaluation sets at a 20\,ms temporal resolution.}
\label{tab:boundary_fusion}
\end{table}

To evaluate whether combining multiple boundary detectors improves
boundary localization accuracy, we perform score-level fusion of
boundary detectors trained with different feature extractors and
augmentation strategies. The resulting performance on the PartialSpoof
and HAD evaluation sets is shown in Table~\ref{tab:boundary_fusion}.

The fused boundary detector achieves a frame-level EER of 2.49\% on
PartialSpoof, substantially improving over the best individual boundary
detector. This result suggests that boundary detection benefits from
model diversity, as different feature representations and augmentation
schemes capture complementary temporal transition cues. Importantly,
since boundary detection errors directly affect segment extraction,
improving boundary localization accuracy has a cascading effect on the
overall performance of the full split-and-conquer pipeline.

\subsection{Step 2: Segment-Level Classification}

\begin{table}[h]
\centering
\small
\footnotesize
\setlength{\tabcolsep}{5pt}
\begin{tabular}{lcccc|ccc|ccc|ccc}
\toprule
\multirow{3}{*}{Feature ext.} & \multirow{3}{*}{Aug.} & \multicolumn{6}{c}{PartialSpoof S-EER (\%)} & \multicolumn{6}{|c}{HAD S-EER (\%)} \\
& & \multicolumn{3}{c}{Dev} & \multicolumn{3}{|c}{Eval} & \multicolumn{3}{|c}{Dev} & \multicolumn{3}{|c}{Eval} \\
& & 1\,s & 2\,s & 4\,s & 1\,s & 2\,s & 4\,s & 1\,s & 2\,s & 4\,s & 1\,s & 2\,s & 4\,s \\
\midrule
\multirow{3}{*}{XLSR128} 
  & None & 1.23 & 1.17 & 1.17 & 5.07 & 4.08 & 3.94 & 0.01 & 0.0 & 0.0 & 0.37 & \textbf{0.02} & \textbf{0.0} \\
  & MS   & 1.26 & \textbf{1.16} & \textbf{1.16} & \textbf{4.31} & \textbf{3.75} & \textbf{3.28} & 0.01 & 0.0 & 0.0 & \textbf{0.01} & 0.64 & \textbf{0.0} \\
  & MF   & \textbf{1.19} & 1.26 & 1.23 & 5.37 & 4.54 & 4.22 & 0.01 & 0.0 & 0.01 & 0.33 & 0.41 & 0.94 \\
\midrule
\multirow{3}{*}{XLSR53} 
  & None & 1.40 & 1.48 & 1.54 & 4.60 & 3.76 & 3.88 & 0.02 & 0.01 & 0.0 & 0.76  & 0.42 & 0.02 \\
  & MS   & 1.37 & 1.42 & 1.50 & 4.72 & 4.14 & 3.78  & 0.02 & 0.0 & 0.01 & 0.03  & 0.46 & 0.04 \\
  & MF   & 1.37 & 1.37 & 1.6 & 5.06 & 3.98 & 3.82 & 0.02 & 0.0 & 0.03 & 0.46 & 0.46 & 0.99 \\

\bottomrule
\end{tabular}
\caption{Segment-level EER (S-EER) on the PartialSpoof and HAD datasets for different training configurations, evaluated on the development and evaluation sets using ground-truth segment boundaries. Results are reported for segment classifiers trained with fixed input durations of 1\,s, 2\,s, and 4\,s, reflecting different target segment lengths employed during segment-level classification.}

\label{tab:segment_eer}
\end{table}

We investigate the intrinsic capability of the segment-level
classifier under ideal segmentation conditions by using ground-truth
boundaries. The experiments evaluate the effect of feature extractors,
augmentation schemes, and fixed input durations on classification
performance. This analysis aims to determine how temporal context and
training configuration influence generalization, revealing that longer
input durations and multi-length modeling provide complementary
representations that can be effectively combined through fusion.
The resulting performance across configurations is summarized in Table~\ref{tab:segment_eer}. On the development set, all configurations achieve similarly low error rates, with only modest variation across input lengths and augmentations. In contrast, clearer trends emerge on the evaluation set: for both XLSR128 and XLSR53, performance consistently improves as the fixed input duration increases from 1\,s to 4\,s, while MaskedSpec systematically outperforms both the no-augmentation and MaskedFeature variants. Notably, most ground-truth segments in PartialSpoof are shorter than 1\,s, meaning that longer inputs are created through reflection rather than by introducing additional speech content. The resulting performance differences therefore indicate that varying the fixed input duration alters the feature-space representation of the same underlying signal, effectively producing distinct views of identical segments. This observation motivates the subsequent use of fusion across input lengths, which exploits these representation differences to improve performance.

The consistent improvement with longer input durations further suggests that segment-level classification benefits from increased temporal context even under ideal ground-truth segmentation. This behavior can be interpreted from two perspectives. First, the self-supervised feature extractors were originally pretrained using substantially longer inputs, typically full utterances or multi-second recordings, whereas the segments in this task may be considerably shorter. Maintaining input durations closer to those encountered during pretraining likely preserves temporal statistics expected by the encoder, enabling more effective feature extraction. Second, extending segments through reflection introduces temporal redundancy that allows the model to aggregate repeated evidence of the same acoustic characteristics. Although no new information is added, repeated contextual exposure may stabilize internal representations and reduce variability in the classifier’s decision space, leading to improved separability between bona fide and spoof segments.

\begin{table}[h]
\centering
\footnotesize
\begin{tabular}{lcc}
\toprule
Fusion configuration & DEV S-EER (\%) & EVAL S-EER (\%) \\
\midrule
PartialSpoof, XLSR128, None, all input fix-lengths & 1.08 & 3.78 \\
PartialSpoof, All & 0.91 & 2.78  \\
HAD, All & 0.0 & 0.01  \\

\bottomrule
\end{tabular}
\caption{Segment-level EER (S-EER) of different fusion configurations on the PartialSpoof and HAD datasets. Results are reported on the development and evaluation sets using ground-truth segment boundaries, isolating the intrinsic performance of the segment classification stage from boundary estimation errors.}

\label{tab:seg_class_fusion}
\end{table}

To assess the effect of score-level fusion at the segment classification
stage, we fuse segment-level scores computed using ground-truth segment
boundaries, thereby isolating the intrinsic performance of the segment
classifiers from boundary estimation errors. The results are shown in
Table~\ref{tab:seg_class_fusion}.
Several observations can be made from these results. First, fusing
segment classifiers trained with different fixed input lengths already
provides a clear improvement over individual models, indicating that
models trained with different temporal contexts capture complementary
segment-level information. This multi-length fusion reduces the
evaluation EER to 3.78\% and the development EER to 1.08\%, demonstrating
that temporal context diversity alone is beneficial for segment-level
classification.

Extending the fusion to include all models presented in
Table~\ref{tab:segment_eer} yields further improvements, reaching
0.91\% EER on the development set and 2.78\% on the evaluation set. This
result suggests that diversity not only in temporal context but also in
feature representation and augmentation strategy contributes to improved
segment classification performance. Overall, these results indicate that
segment-level classification benefits significantly from model diversity,
and that score-level fusion effectively leverages complementary decision
patterns across different model configurations.

\subsection{Full Pipeline}

\begin{table}[h]
\centering
\begin{tabular}{lc|ccc|ccc}
\toprule
\multirow{3}{*}{Feature extctor} & \multirow{3}{*}{Augmentation} & \multicolumn{6}{c}{EVAL F-EER (\%)} \\ 
& & \multicolumn{3}{c}{PartialSpoof} & \multicolumn{3}{|c}{HAD} \\

& & 1\,s & 2\,s & 4\,s & 1\,s & 2\,s & 4\,s \\
\midrule
\multirow{3}{*}{XLSR128} 
  & None & 10.92 & 8.70 & 8.03 & 0.03 & 0.03 & 0.02 \\
  & MS   & 9.26  & 8.18 & 7.65 & 0.15 & 0.06 & 0.02\\
  & MF   & 12.47 & 8.61 & 8.08 & 0.27 & 0.14 & 0.02\\
\midrule
\multirow{3}{*}{XLSR53} 
  & None & 9.38  & 8.75 & 9.15 & 0.04 & 0.12 & 0.15 \\
  & MS   & 9.64  & 9.41 & 9.53 & 0.05 & 0.08 & 0.05\\
  & MF   & 10.66 & 8.90 & 9.13 & 0.06 & 0.04 & 0.10\\
\bottomrule
\end{tabular}
\caption{Frame-level EER (F-EER) of the complete split-and-conquer pipeline on the PartialSpoof and HAD evaluation sets, evaluated at a 20\,ms frame resolution using predicted boundaries. Performance is reported for segment classifiers trained with fixed input durations of 1\,s, 2\,s, and 4\,s, reflecting different target segment lengths used during segment-level classification.}

\label{tab:frame_eer_ps}
\end{table}

We evaluate the full split-and-conquer pipeline, where temporal boundaries are first estimated by the boundary detector and the authenticity of the resulting segments is subsequently determined by the segment-level classifier. In contrast to the previous evaluations, this experiment assesses the complete system with all components operating jointly, enabling us to quantify the end-to-end impact of boundary estimation errors on the final detection performance. In particular, this experiment examines whether the trends observed under ideal segmentation conditions remain valid when segmentation is imperfect and must be inferred automatically.

Table~\ref{tab:frame_eer_ps} reports frame-level EER results for the complete pipeline on the PartialSpoof and HAD evaluation sets, where segment boundaries are predicted by the boundary detector and evaluation is performed at a 20,ms frame resolution. Several trends emerge from these results. For XLSR128-based models, performance improves as the fixed input duration increases from 1,s to 4,s, indicating that longer temporal context produces more reliable segment-level representations and reduces sensitivity to boundary localization errors. MaskedSpec augmentation consistently provides the strongest performance for XLSR128 across segment durations, with the XLSR128+MS+4,s configuration achieving the lowest frame-level EER of 7.65\% on PartialSpoof. Overall, these results suggest that robustness to imperfect segmentation is influenced both by the temporal context used for segment classification and by augmentation strategies that improve feature generalization.

Performance differences between configurations become more pronounced when predicted boundaries are used instead of ground-truth segmentation, indicating that segment classifiers differ not only in classification accuracy but also in their robustness to boundary localization errors. Consequently, the full pipeline evaluation reveals robustness characteristics that are not observable under ideal segmentation conditions. These results highlight that robustness to imperfect segmentation is a key factor in partial spoof detection systems, and that improvements in segment-level classification accuracy do not necessarily translate equally to end-to-end performance when boundary estimation errors are present.

\begin{table}[h]
\centering
\begin{tabular}{lcc}
\toprule
\multirow{2}{*}{Fusion} & \multicolumn{2}{c}{EVAL F-EER (\%)} \\
 & GT boundaries & Detected boundaries \\
\midrule
PartialSpoof, All & 2.78 & 6.55 \\
HAD, All          & 0.01 & 0.01 \\
\bottomrule
\end{tabular}
\caption{Fusion performance of the full pipeline under ground-truth (GT) versus detected boundaries, on PartialSpoof and HAD.}
\label{tab:full_fusion}
\end{table}

To assess the performance of the fused system and the impact of boundary
estimation errors in the complete pipeline, we evaluate score-level
fusion under both ground-truth (GT) and detected boundaries on the
PartialSpoof and HAD evaluation sets. The fused system combines the
single systems listed in Table~\ref{tab:segment_eer} and
Table~\ref{tab:frame_eer_ps}, respectively, using score-level fusion,
where the final decision score is computed as the average of the
individual system scores. The results are presented in
Table~\ref{tab:full_fusion}.

On PartialSpoof, the fusion system achieves an EER of 2.78\% when
ground-truth boundaries are used and 6.55\% when boundaries are predicted
by the boundary detector. The gap between these two conditions quantifies
the performance degradation attributable to boundary estimation errors,
since the segment classifiers and fusion method remain identical.
Nevertheless, even under detected boundaries, the fused system maintains
substantially lower error rates than individual systems, demonstrating
that fusion improves robustness not only to spoofing variability but also
to segmentation errors.

On HAD, both ground-truth and detected boundary conditions result in the
same EER of 0.01\%, indicating that boundary estimation errors have
negligible impact on this dataset. This behavior suggests that spoof regions in HAD are either easier to localize or less sensitive to small boundary misalignments, and that segment-level classification dominates overall performance in this dataset. A likely explanation is that spoofed segments in HAD are generally longer than those in PartialSpoof, making the system less
sensitive to boundary localization errors, as small boundary
misalignments affect only a small portion of the segment while most of
the segment content remains correctly labeled. In addition, boundary
detection accuracy on HAD was higher overall, further reducing the
impact of segmentation errors on the final detection performance.

\begin{figure}[h!]
\centering
\includegraphics[width=0.8\textwidth]{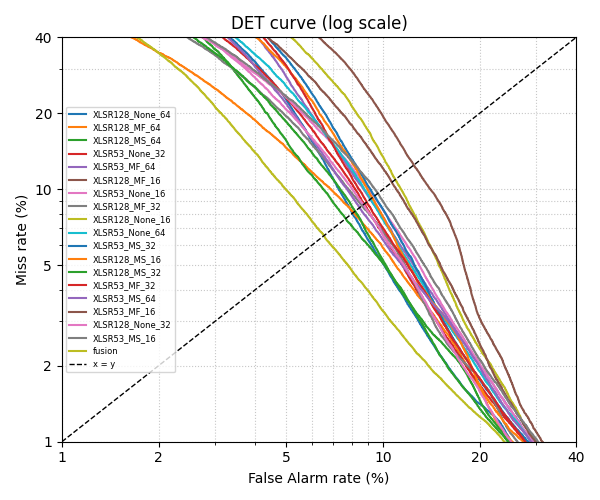} 
\caption{
Detection error trade-off (DET) curves of all single models and their score-level fusion on the
evaluation set. Each curve reports the miss rate as a function of the false alarm rate, with both axes
shown up to 40\%. Individual configurations in the legend are denoted using the format
\textit{feature extractor\_augmentation\_fixed input length}, describing the feature representation,
training augmentation, and segment input duration used by each model. While individual models exhibit
complementary strengths in different operating regions, the fused system consistently dominates the
single-model baselines, achieving lower miss rates across most false alarm rates and yielding the best
overall trade-off between bona fide rejection and spoof acceptance.}
\label{det_curve}
\end{figure}

Further insight into the behavior of individual systems and their fusion
across operating points is provided in Figure~\ref{det_curve}. .
The DET curves show that individual models exhibit different trade-offs
between miss and false alarm rates, indicating that different
configurations perform better in different operating regions. The fused
system, however, consistently dominates the single-system baselines
across most operating points, achieving lower miss rates for the same
false alarm rates. This behavior indicates that the individual systems
provide complementary information, and that score-level fusion
effectively reduces both miss and false alarm rates by averaging out
individual model errors. Overall, these results demonstrate that fusion
not only improves absolute performance but also increases system
stability and robustness in the full split-and-conquer pipeline.
In the figure legend, each system is denoted using the format
\textit{feature extractor\_augmentation\_fixed input length}, which
indicates the feature representation used by the model, the augmentation
strategy applied during training, and the fixed segment input duration
used for segment-level classification. This notation allows direct
comparison between systems that differ in feature extractor,
augmentation method, or temporal context

\begin{table}[h!]
\centering
\begin{tabular}{lcccc}
\toprule
\multicolumn{5}{c}{HAD} \\
\midrule
Method & EER $\downarrow$ (\%) & Precision $\uparrow$ (\%) & Recall $\uparrow$ (\%) & F1 $\uparrow$ (\%) \\
\midrule
MRM \cite{zhang2022partialspoof}   & 0.18 & 99.96 & 99.82 & 99.89 \\
IFBDN \cite{Cai2024BoundaryDeepfake} & 0.35 & 99.92 & 99.65 & 99.78 \\
CFPRF \cite{Wu2024CFPRF}           & 0.08 & 99.98 & 99.92 & 99.95 \\
Split\&Conquer (Ours) & \textbf{0.01} & \textbf{99.99} & \textbf{99.99} & \textbf{99.99}\\
\midrule
\multicolumn{5}{c}{PartialSpoof} \\
\midrule
Method & EER $\downarrow$ (\%) & Precision $\uparrow$ (\%) & Recall $\uparrow$ (\%) & F1 $\uparrow$ (\%) \\
\midrule
MRM   & 13.72 & 80.20 & 86.46 & 83.21 \\
IFBDN &  9.68 & 93.72 & 90.32 & 91.99 \\
CFPRF &  7.61 & 88.77 & 92.39 & 90.54 \\
Split\&Conquer (Ours)  & \textbf{6.55} & \textbf{95.75} & \textbf{93.44} & \textbf{94.58} \\
\bottomrule
\end{tabular}
\caption{Frame-level performance comparison on the HAD and PartialSpoof datasets with recently published methods. Evaluation is conducted at a 20\,ms temporal resolution using predicted boundaries, and results are reported in terms of EER, precision, recall, and F1 score.}

\label{tab:prior__}
\end{table}

To evaluate the effectiveness of the proposed split-and-conquer framework
in relation to existing partial spoof detection methods, we compare our
approach with several recently published methods on the HAD and
PartialSpoof datasets. The comparison focuses on frame-level performance
under predicted boundary conditions, which reflects the complete pipeline
evaluation scenario. The results are summarized in
Table~\ref{tab:prior__}.

On PartialSpoof, the proposed method achieves the lowest EER and the
highest precision, recall, and F1 score among all compared methods,
indicating improved detection accuracy as well as a better balance
between false alarms and missed detections. The improvement in both EER
and F1 suggests that the proposed boundary-first formulation and
segment-level classification strategy provide more reliable frame-level
decisions compared to prior approaches that rely primarily on
frame-level labeling.

On HAD, all methods achieve very low error rates, indicating that this
dataset is generally less challenging for partial spoof detection.
Nevertheless, the proposed method still achieves the best overall
performance across all reported metrics. The smaller performance gap
between methods on HAD compared to PartialSpoof suggests that the main
differences between approaches become more evident on more challenging
datasets, where accurate boundary localization and robust segment-level
classification play a larger role in overall system performance.

\begin{table}[h]
\centering
\begin{tabular}{lccccccc}
\toprule
Method & 0.02\,s & 0.04\,s & 0.08\,s & 0.16\,s & 0.32\,s & 0.64\,s & Utt. \\
\midrule
MRM   & 13.72 & 14.46 & 15.29 & 11.60 &  9.63 &  7.24 & 1.48 \\
CFPRF &  7.61 &  7.36 &  6.84 &  6.04 &  5.24 &  4.80 & 1.72 \\
Split\&Conquer (Ours)  & \textbf{6.55} & \textbf{6.30} & \textbf{5.90} & \textbf{5.29} & \textbf{4.53} & \textbf{3.86} & \textbf{0.58} \\
\bottomrule
\end{tabular}
\caption{Frame-level EER (\%) at different temporal resolutions on PartialSpoof.}
\label{tab:resolution_ps}
\end{table}

To analyze how detection performance varies with temporal resolution
and to evaluate whether the proposed method maintains its advantage
under coarser temporal evaluation conditions, we compare frame-level EER
across multiple temporal resolutions on the PartialSpoof dataset. The
results are presented in Table~\ref{tab:resolution_ps}.

Several observations can be made from these results. First, all methods
show improved EER as the temporal resolution becomes coarser, which is
expected since temporal smoothing reduces the impact of short boundary
misalignments and isolated frame-level errors. The proposed
split-and-conquer method consistently achieves the lowest EER across all
resolutions, indicating that the performance gains are not limited to a
specific evaluation granularity. In particular, the performance gap
between the proposed method and prior work remains consistent as the
evaluation window increases, suggesting that the improvements stem from
better segment-level modeling rather than only more precise boundary
localization.

This behavior is particularly relevant for practical deployment
scenarios, where downstream applications often operate at coarser
temporal scales, such as word-level or phrase-level decisions, and where
small temporal misalignments are acceptable. The consistent advantage
across resolutions therefore indicates that the proposed method
improves overall detection quality and robustness, rather than only
optimizing performance for a specific frame-level evaluation protocol.

\section{Analysis}

\paragraph{PartialSpoof per-utterance view} 

\begin{figure}[h!]
\centering
\includegraphics[width=0.7\textwidth]{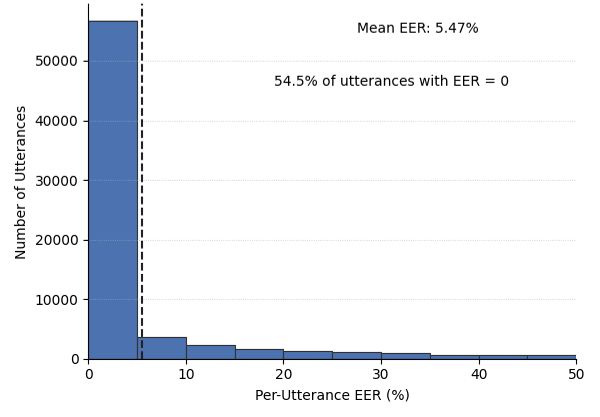} 
\caption{\it Distribution of per-utterance EER obtained using the complete pipeline on the PartialSpoof evaluation set (71,239 utterances). Each bin represents the EER computed independently for a single utterance. The dashed vertical line indicates the average EER of 5.47\%. Notably, 54.5\% of utterances achieve zero EER, highlighting the skewed performance distribution across samples.}
\label{fig:eers}
\end{figure}

The per-utterance EER distribution in Figure~\ref{fig:eers} is highly skewed, with more than half of the utterances (54.5\%) achieving an EER of zero and most remaining examples concentrated at relatively low error rates, while only a small fraction form a long tail of difficult cases. The mean per utterance EER is 5.47\%, which is lower than the global frame level EER of 6.55\%. This difference reflects the fact that the frame level metric weights all frames equally, so a few long and challenging utterances with localized errors can disproportionately influence the overall EER, whereas the utterance averaged metric assigns equal weight to each utterance and better captures the typical performance across the corpus.

An analysis of the utterances associated with high EER values did not reveal consistent patterns or common characteristics based on the metadata and information provided with the dataset. This suggests that the difficulty of these cases may not be explained solely by obvious acoustic conditions or dataset attributes. Investigating whether there exist underlying criteria that make certain spoofing attacks particularly challenging, beyond simple distribution mismatch with the training data, remains an interesting direction for future work.

\paragraph{Dataset analysis}

\begin{figure}[h!]
\centering
\includegraphics[width=0.8\textwidth]{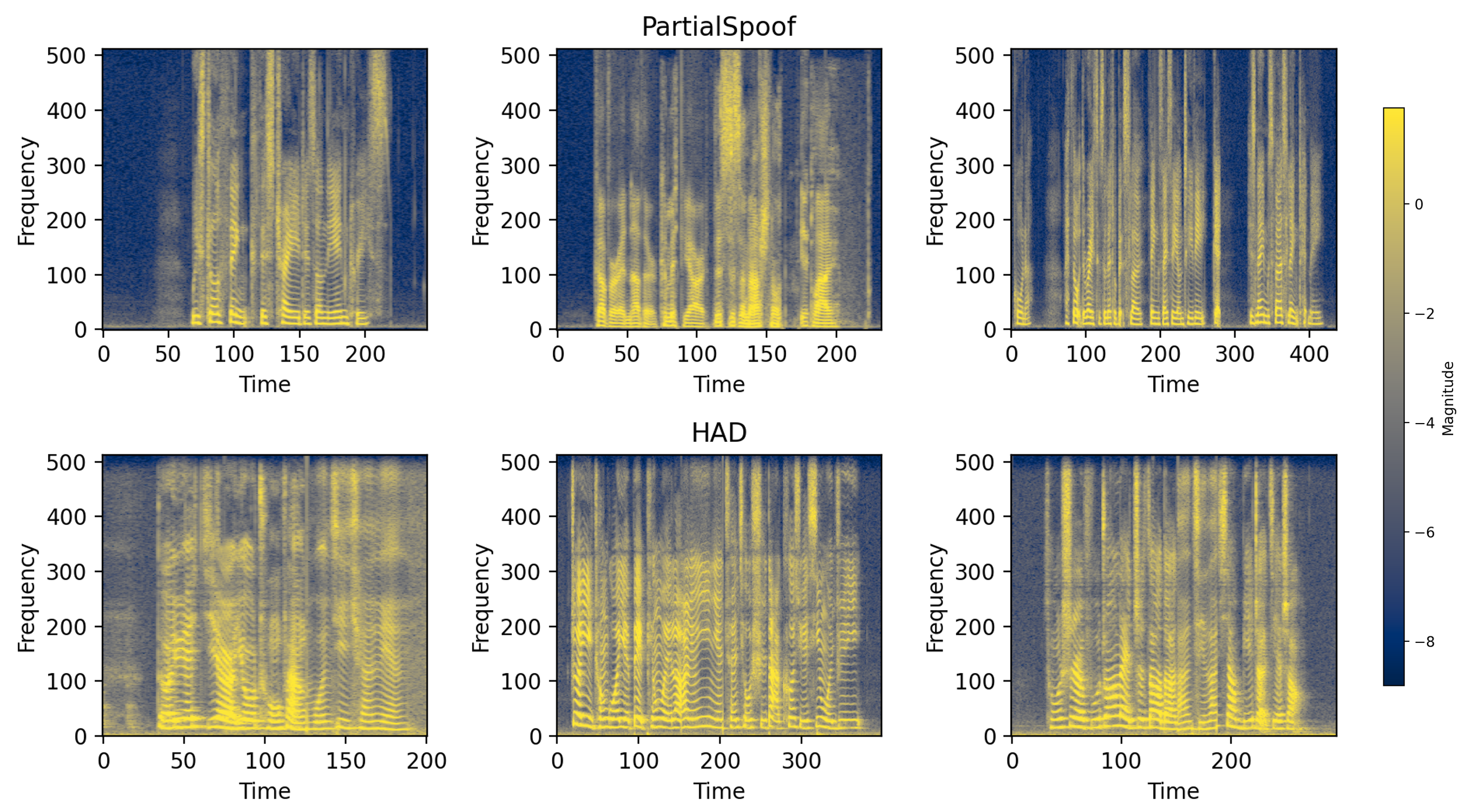} 
\caption{\it Log-magnitude spectrogram examples from three corpora. Top row: PartialSpoof, English. Middle row: HAD, Mandarin. Bottom row: LPS, English.}
\label{fig:specs}
\end{figure}

To better understand the acoustic differences between the PartialSpoof
and HAD datasets and their potential impact on partial spoof detection,
we examine log-magnitude spectrograms of randomly selected utterances
from both datasets, shown in Figure~\ref{fig:specs}. Even a coarse visual
inspection reveals pronounced corpus-specific differences. In the HAD
(Mandarin) examples, the spectrograms exhibit relatively dense and continuous spectral structure with fewer extended low-energy or silent regions, indicating more tightly packed syllables and smoother transitions between them. This pattern is consistent with contrastive phonetic studies showing that Mandarin, as a tonal language with a rich inventory of initial–final combinations, makes extensive use of connected-speech processes such as assimilation, weakening, and tone sandhi in continuous speech \cite{liao2019_phonetic_change}.

In contrast, the PartialSpoof utterances show more moderate spectral occupancy and shorter durations, with clearer alternation between speech and low-energy regions, reflecting their construction from shorter English sentences with localized manipulations. Overall, these observations highlight substantial differences in language, prosody, and recording characteristics between the two corpora.

\paragraph{Temporal Forgery Localization Analysis}

\begin{table*}[h]
\centering
\footnotesize
\setlength{\tabcolsep}{1.5pt}
\begin{tabular}{lcccccccccc}
\toprule
Method & AP@0.5 & AP@0.75 & AP@0.9 & AP@0.95 & mAP & AR@1 & AR@2 & AR@5 & AR@10 & AR@20 \\
\midrule
MRM      & 54.25 & 46.47 & 40.57 & 36.70 & 46.68 & 19.70 & 36.10 & 58.09 & 65.40 & 66.51 \\

IFBDN     & 58.65 & 49.30 & 41.39 & 35.33 & 48.79 & 18.52 & 34.77 & 55.41 & 64.47 & 62.23 \\

CFPRF         & 66.34 & 55.47 & 48.05 & 40.96 & 55.22 & 18.48 & 35.57 & 58.06 & 65.47 & 66.53 \\

\textbf{Split\&Conquer} & \textbf{78.51} & \textbf{72.06} & \textbf{63.93} & \textbf{56.72} & \textbf{71.27} & \textbf{28.19} & \textbf{48.81} & \textbf{76.41} & \textbf{85.85} & \textbf{87.42} \\
\bottomrule
\end{tabular}
\caption{Temporal forgery localization comparison on the PartialSpoof (PS)
dataset. Prior results are taken from \cite{Wu2024CFPRF}. Our method
reports AP and AR using the same evaluation protocol.}
\label{tab:temporal_localization}
\end{table*}


To further evaluate temporal forgery localization performance, we compare
our method with prior work reported in \cite{Wu2024CFPRF} on the
PartialSpoof (PS) dataset. Following their evaluation protocol,
performance is measured using Average Precision (AP) at multiple IoU
thresholds and Average Recall (AR) under different proposal budgets.

Table~\ref{tab:temporal_localization} presents a comparison of temporal
forgery localization performance on the PartialSpoof dataset. The proposed
method consistently outperforms prior approaches across all evaluated
metrics. In particular, the substantial improvements at higher IoU
thresholds indicate that predicted segments align more precisely with the
ground-truth manipulation regions, suggesting improved temporal boundary
accuracy rather than merely increased overlap. The strong gains in mAP
further demonstrate stable localization performance across varying overlap
requirements, indicating that improvements are not confined to a specific
operating point.

In addition to precision, the method achieves markedly higher recall across
all proposal budgets. The rapid increase in AR values with a small number of
allowed proposals suggests that forged regions are identified efficiently,
with fewer redundant candidates required to achieve high coverage. This
behavior indicates that the predicted segments are both well localized and
consistently positioned around manipulated regions. Overall, the results
show that improved boundary estimation translates into tighter temporal
localization and more reliable detection of forged speech segments.

\section{Conclusion}

This work introduced a split-and-conquer framework for partial deepfake
speech detection that decomposes the task into boundary detection
followed by segment-level classification. By separating temporal
localization from authenticity classification, the proposed approach
simplifies the learning objective and allows each component to focus on
a well-defined subproblem. The boundary detector identifies transition
points between acoustic regions, enabling the signal to be partitioned
into candidate spoof-uniform segments that are subsequently evaluated
independently.

Experimental results demonstrate that this formulation leads to
consistent improvements in both detection and localization performance.
Boundary detection exhibits stable behavior across feature
representations, suggesting that transition cues constitute a reliable
signal. At the segment level, performance benefits from increased
temporal context even when segments are artificially extended,
indicating the importance of temporal statistics learned during
self-supervised pretraining. When integrated into the full pipeline,
these properties translate into improved frame-level detection and
substantially stronger temporal localization results, particularly at
strict overlap criteria, showing more precise alignment with manipulated
regions.

Beyond performance gains, the analysis provides several insights into
partial spoof detection. First, accurate boundary estimation reduces the
complexity of the classification task by operating on acoustically
consistent segments. Second, different input durations induce distinct
feature-space representations, motivating fusion strategies that improve
robustness. Finally, localization results indicate that precise temporal
segmentation can achieve strong coverage with relatively few candidate
regions, suggesting an efficient alternative to proposal-heavy
localization pipelines.

Despite these advantages, several limitations remain. Boundary
prediction errors may propagate to later stages and affect final
decisions, particularly in highly ambiguous transition regions.
Additionally, the approach assumes that manipulated content can be
approximated by piecewise-uniform segments, which may not fully capture
gradual or highly subtle manipulations. Future work will explore joint
optimization of boundary detection and classification, improved
adaptation to domain shifts, and more flexible modeling of temporal
structure.

Overall, the proposed framework demonstrates that simplifying the
learning objective through explicit temporal decomposition provides an
effective and interpretable solution for partial deepfake speech
detection and localization.

\bibliographystyle{IEEEtran}
\bibliography{references}
\end{document}